\documentclass[12pt]{article}
\usepackage{graphics}
\usepackage{amssymb}
\usepackage{amsmath}

\newcommand{\rf}[1]{(\ref{#1})}
\newcommand{\beq}{\begin{equation}}
\newcommand{\eeq}{\end{equation}}
\newcommand{\bea}{\begin{eqnarray}}
\newcommand{\eea}{\end{eqnarray}}

\newcommand{\e}{\mbox{e}}

\renewcommand{\l}{\lambda}
\newcommand{\La}{\Lambda}

\newcommand{\Del}{\Delta}

\newcommand{\oh}{\frac{1}{2}}

\newcommand{\dg}{\dagger}

\newcommand{\tr}{\mathrm{tr}\,}
\newcommand{\ra}{\rangle}
\newcommand{\la}{\langle}
\newcommand{\prt}{\partial}

\newcommand{\cN}{{\cal N}}

\newcommand{\tg}{{\tilde{g}}}

\newcommand{\hI}{{\hat{I}}}

\newcommand{\bt}{{\bar{t}}}
\newcommand{\bg}{{\bar{g}}}
\newcommand{\bz}{{\bar{z}}}
\newcommand{\bV}{{\bar{V}}}

\newcommand{\cdtL}{\La_{{\rm cdt}}}
\newcommand{\cdtZ}{Z_{\rm cdt}}
\newcommand{\cdtW}{W_{\rm cdt}}

\begin{document}

\begin{center}

{ \large \bf A new continuum limit of matrix models }

\vspace{30pt}

{\sl J.\ Ambj\o rn}$\,^{a,b}$, {\sl R.\ Loll}$\,^{b}$,
{\sl Y.\ Watabiki}$\,^{c}$, {\sl W.\ Westra}$\,^{d}$ and
{\sl S.\ Zohren}$\,^{e,f}$

\vspace{24pt}

{\footnotesize

$^a$~The Niels Bohr Institute, Copenhagen University\\
Blegdamsvej 17, DK-2100 Copenhagen \O , Denmark.\\
{ email: ambjorn@nbi.dk}\\

\vspace{10pt}

$^b$~Institute for Theoretical Physics, Utrecht University, \\
Leuvenlaan 4, NL-3584 CE Utrecht, The Netherlands.\\
{ email: loll@phys.uu.nl}\\

\vspace{10pt}

$^c$~Tokyo Institute of Technology,\\ 
Dept. of Physics, High Energy Theory Group,\\ 
2-12-1 Oh-okayama, Meguro-ku, Tokyo 152-8551, Japan\\
{email: watabiki@th.phys.titech.ac.jp}\\

\vspace{10pt}

$^d$~Department of Physics, University of Iceland,\\
Dunhaga 3, 107 Reykjavik, Iceland\\
{ email: wwestra@raunvis.hi.is}\\

\vspace{10pt}

$^e$~Mathematical Institute, Leiden University,\\
Niels Bohrweg 1, 2333 CA Leiden, The Netherlands\\
{email: zohren@math.leidenuniv.nl}\\

\vspace{10pt}

$^f$~Blackett Laboratory, Imperial College,\\
London SW7 2AZ, UK, and\\
{email: stefan.zohren@imperial.ac.uk}

}
\vspace{48pt}

\end{center}


\begin{center}
{\bf Abstract}
\end{center}

We define a new scaling limit of matrix models which
can be related to the method of causal dynamical 
triangulations (CDT) used when investigating two-dimensional
quantum gravity. Surprisingly, the new scaling limit of
the matrix models is also a  matrix model, thus 
explaining why the recently developed CDT continuum string field theory
(arXiv:0802.0719) has a matrix-model representation (arXiv:0804.0252).

\vspace{12pt}
\noindent


\newpage

\section{Introduction}\label{sec1}

The great versatility of matrix models or matrix integrals in theoretical
physics is well illustrated by their particularly 
beautiful application in two-dimensional 
Euclidean quantum gravity (see \cite{david1,gm,difgz,book}  
for reviews). This theory 
can be defined as a suitable sum over triangulations, so-called 
``dynamical triangulations'' (DT), whose continuum
limit is obtained by taking the side lengths $a$ 
of the triangles to zero. 
The method of DT was originally introduced as a nonperturbative worldsheet 
regularization of the Polyakov bosonic string \cite{ambjorn,david,mkk}.
There it was used with success 
(or to disappointment, depending on one's taste) 
to show rigorously that a tachyon-free version
of Polyakov's bosonic string theory does not exist in target 
space dimensions $d > 1$ \cite{ad}. 
However, when viewed as a theory of 2d quantum 
gravity coupled to matter with central charge $c\leq 1$, 
the theory -- noncritical string theory -- is perfectly consistent,
and matrix models have been used to solve
in an elegant way the combinatorial aspects of the 
DT construction, where one sums over random surfaces
glued together from equilateral triangles.

The DT approach possesses a well-defined cut-off, the 
length $a$ of the lattice links. As has been discussed in many
reviews (for instance the ones mentioned above), 
a continuum limit can be defined when the 
lattice spacing is taken to zero while simultaneously renormalizing
the bare cosmological constant and possibly other matter
coupling constants. However, the continuum limit in question has
some unconventional properties. In this article we will
show that there is another way of taking the scaling limit of 
the matrix models, which still relates them to a summation
over triangulated random surfaces, the so-called 
{\it causal dynamical triangulations} (CDT) \cite{al}. We will
show that this limit is in a way more natural and does not 
lead to the somewhat unconventional renormalization
encountered in the standard DT approach.

In the next section, we will review briefly the conventional situation 
in the simplest case, that of pure Euclidean two-dimensional 
quantum gravity without matter. We follow the notations in \cite{book}. 
In Sec.\ \ref{sec3} we introduce a new limit
of the corresponding matrix model and analyze its interpretation
in terms of random surfaces. Sec.\ \ref{discuss}  
discusses the results obtained and outlines possible applications.

\section{The old matrix model and its continuum limit}\label{sec2}

Our starting point is the Hermitian matrix integral 
\beq\label{2.1}
Z(\tilde{g}) = \int d\phi \; 
\e^{-N \tr \left(\oh \phi^2 -\frac{\tg}{3} \phi^3\right)}
= \sum_{k=0}^\infty \frac{1}{ k!}\;
\int d\phi \; \e^{-\oh N \tr \left( \phi^2\right)}\; 
 \left( \frac{N\tg}{3} \tr \phi^3\right)^k, 
\eeq 
where $\phi$ is an $N\times N$ Hermitian matrix.
This integral is formal since (assuming $\tilde{g}>0$) 
it is not convergent unless an
analytic continuation is performed. However, by power-expanding 
$\exp( N \tg\; \tr \phi^3)$ as indicated in eq.\ \rf{2.1} the matrix integral
defines a formal power series in the coupling constant $\tg$. 
Furthermore, this expansion
can be given the geometric interpretation of gluing together triangles
in all possible ways. The size $N$ of the matrix acts as a factor, organizing 
the power series
further into a summation over surfaces of fixed topology.\footnote{By 
expanding the matrix integral in a power series in $\tg$ and subsequently
in powers of $1/N^2$, we are of course leaving out any nonperturbative 
contributions
to the matrix integral (defined in some way, e.g.\ by analytic continuation)
which are not captured by such an expansion.} It turns
out that the power series corresponding to a given fixed topology
is convergent with convergence radius $\tg_c$ independent of the topology 
considered, a property that has made the matrix integrals 
useful in the study of noncritical strings.
The particular rearrangement of the power series \rf{2.1} according to
topology, i.e.\ in powers of $1/N^2$, is called the large-$N$ expansion.

Let us briefly review how a ``continuum limit'' of the matrix integrals
can be associated with noncritical string theory in the simplest case
of $c=0$, ``pure'' Euclidean two-dimensional quantum gravity,
where there is no extended target space (equivalently, 
no matter fields coupled to 2d gravity).
This is best done by calculating an ``observable'', the so-called
disc amplitude or Hartle-Hawking wave function $w(z)$ of the 2d universe to 
leading order in the $1/N$ expansion. One has 
\beq\label{2.2}
w(z) \equiv \left\langle \frac{1}{N}\;
\tr \frac{1}{z-\phi} \right\rangle = 
\frac{1}{N}\;\sum_{n=0}^\infty \frac{\la \tr \phi^n \ra}{z^{n+1}}, 
\eeq
where 
\beq\label{2.3}
\la \tr \phi^n \ra \equiv Z(\tg)^{-1} \int d\phi \; \tr \phi^n \; 
\e^{-N \tr \left(\oh \phi^2 -\frac{\tg}{3} \phi^3\right)},
\eeq
again to be viewed as a formal power series in $\tg$.
To leading order in $1/N$ one finds
\beq\label{2.4}
w(z) = \oh \left(z- \tg z^2 +
\tg(z-b) \sqrt{(z-c)(z-d)}\right),~~~c>0,~~c \ge |d|,
\eeq
where the constants $b(\tg)$, $c(\tg)$ and $d(\tg)$ 
are functions of $\tg$ and uniquely determined by the 
requirement that $w(z)$ fall off like $1/z$ 
(with coefficient 1) for $z \to \infty$.

\begin{figure}
\centerline{\scalebox{0.8}{\rotatebox{0}{\includegraphics{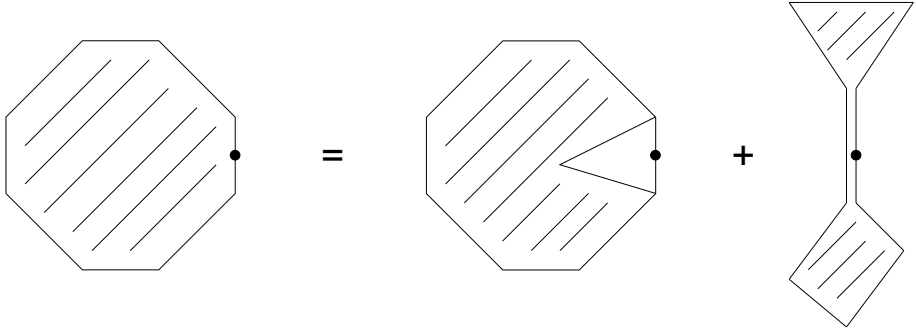}}}}
\caption{{\small 
Graphical representation of relation \rf{2.4a}: The boundary contains 
one marked link which is part of a triangle or a double link. 
Associated to each
triangle is a weight $\tilde g$, and to each double link a weight 1. 
}}
\label{fig1}
\end{figure}

The geometric interpretation of $w(z)$ is as follows: the term in
\rf{2.2} corresponding to $\la \tr \phi^n \ra$ represents the summation
over all triangulations with a boundary consisting of $n$ links and
(to leading order in the large-$N$ expansion) the topology
of a disc.\footnote{When performing the Gaussian integral, each boundary 
link, represented by a factor of $\phi$ in $\tr \phi^n$, is glued either to
a triangle or to another boundary link. In the latter case 
we call it a ``double link'' and it will not be glued to
other triangles, see Fig.\ \ref{fig1}.}
Consequently, $ z w(z)$ represents the sum over all triangulations
with the topology of a disc and the constant $\ln z$ has the interpretation
of a boundary cosmological constant $\l_b$, such that a boundary of length
$n$ is assigned a weight $e^{-\l_b n}$. In the same way as $w(z)$ has
a power expansion in $\tg$, it also has a power expansion in $1/z$ and the 
radius of convergence is $c(\tg)$. The geometric interpretation can 
be further elaborated upon by noting that $w(z)$ satisfies the
combinatorial equation 
\beq\label{2.4a}
w(z) = z\tg\, w(z) +\frac{1}{z}\, w^2(z) +\frac{1}{z}\, Q(z,\tg),
\eeq
where $Q(z,\tg)$ is a polynomial in $z$ which is uniquely 
determined\footnote{For a combinatorial interpretation of
$Q(z,\tg)$ see, for instance, \cite{book}.  $Q(z)$ is present in 
eq.\ \rf{2.4a} because Fig.\ \ref{fig1} is  not correct when 
the boundary consists of less than two links.} 
by the requirement that the solution to \rf{2.4a} have the form \rf{2.4}
and fall off like $1/z$.
The recursion relation leading to \rf{2.4a} 
is represented graphically in Fig.\ \ref{fig1}
and will be important in the next section.

The continuum limit of the 
matrix model, relevant for noncritical string theory, is obtained
as follows. Note first that the radius of convergence $\tilde{g}_c$ is 
determined by the condition
\beq\label{2.5a}
b(\tg)=c(\tg),
\eeq
namely, the point
where $w(z)$ changes analytic structure from $(z-c)^{1/2}$ to $(z-c)^{3/2}$. 
Next we fine-tune $\tg$ to $\tg_c$ and $z$ to $c(\tg_c)$ according to
\beq\label{2.5}
\tg = \tg_c\,(1 - \La a^2 +O(a^4)),~~~~z = c(\tg_c) +a Z+O(a^2).
\eeq
These assignments can be viewed as additive renormalizations of the
bare cosmological and boundary cosmological constants such that 
$\La$ and $Z$ now represent the renormalized
coupling constants in the limit where the lattice spacing $a \to 0$.
The rationale behind interpreting $\La$ as a continuum cosmological 
constant comes from considering the term in $w(z)$ 
with $k$ triangles glued together, for very large $k$. 
It will appear as 
\beq\label{2.6}
s(k)\left(\frac{\tg}{\tg_c}\right)^k \approx s(k)\, \e^{-k a^2 \La},
\eeq
where $s(k)$ is a subleading term and we associate $A(k) = k a^2$
with a macroscopic {\it area}. The coupling constant multiplying this
area is by definition proportional to 
the continuum renormalized cosmological constant.
Similarly one is led to the conclusion that $Z$ can be interpreted
as a continuum boundary cosmological constant. 

Inserting \rf{2.5} into \rf{2.4}, we obtain in the limit of $a \to 0$ that
\beq\label{2.6a}
w(z) = \oh \left(z -\tg z^2 + \tg \sqrt{c(\tg_c)-d(\tg_c)}\; 
a^{3/2} W_E(Z,\La)\Big(1+ O(a)\Big)\right),
\eeq
where
\beq\label{2.6b}
W_E(Z,\La) = (Z- \sqrt{2\La/3}) \sqrt{Z+2\sqrt{2\La/3}}
\eeq
is called the continuum Hartle-Hawking wave function or (using
string terminology) the continuum disc amplitude.

\begin{figure}
\centerline{\scalebox{0.8}{\rotatebox{0}{\includegraphics{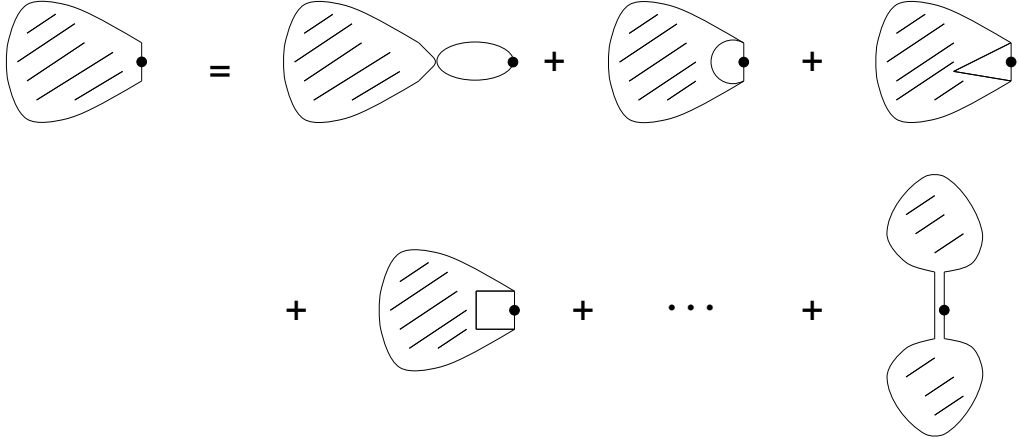}}}}
\caption{{\small 
Graphical representation of relation \rf{2.7a}: the marked link
of the boundary belongs either to an $i$-gon (associated weight $g t_i$) 
or a double link (associated weight 1).
It is also a graphical representation of eq.\ \rf{3.18a}
if instead of weight 1 we associate a weight $g_s$ to the marked double link.
}}
\label{fig2}
\end{figure}

The above arguments can be generalized from the matrix potential
in \rf{2.1} to a general potential of the form
\beq\label{2.7}
V(\phi) =  \oh \,\phi^2 -g \sum_{i=1}^n \frac{t_i}{i} \, \phi^i,
\eeq
again leading to a formal power series after expanding the 
exponential in powers of the coupling constant $g$ and performing the
remaining Gaussian integrals. The geometric interpretation is analogous to
that for the cubic potential, only that now we glue together $i$-gons and $j$-gons, 
where $i$ and $j$ run from 1 to $n$, with 
each $i$-gon assigned a relative weight $t_i$.
By allowing such polygons we are clearly considering a generalization of 
simplicial complexes, but we will for convenience continue to refer to
these as ``triangulations".
The Hartle-Hawking wave function $w(z)$ now satisfies the 
combinatorial equation
\beq\label{2.7a}
w(z)= g\left(\sum_{i=1}^n t_i z^{i-2}\right) w(z) + \frac{1}{z}\, w^2(z) 
+\frac{1}{z}\,Q(z,g),
\eeq
where again the polynomial $Q(z,g)$ is determined by 
the requirement that $w(z)$ fall off like $1/z$ and have 
a single cut. The equation has the graphical representation
shown in Fig.\ \ref{fig2}.

As long as the weights $t_i$ are positive (a natural requirement
if one wants to assign an area to the $i$-gon), one obtains 
\beq\label{2.8}
w(z) = \oh \left(V'(z) + g t_n \sqrt{c(g_c)-d(g_c)}\; a^{3/2} 
W_E(Z,\La)\right)
\eeq
in the limit $a \to 0$. In \rf{2.8}, $V'(z)$ 
denotes the derivative with respect 
to $z$ of the potential $V(z)$ defined in \rf{2.7}, $g_c$
the critical value of $g$ (for fixed $t_i$'s),
i.e.\ the radius of convergence of the formal power expansion in $g$
for fixed topology (in this case that of the disc), and one
has made an ansatz similar to \rf{2.5}, namely, 
\beq\label{2.9}
g = g_c(1-\La a^2),~~~~~z = c(g_c)+a Z
\eeq
for the coupling constants. 
One peculiar aspect of \rf{2.6a} and \rf{2.8} is the non-scaling part
$V'(z)/2$. This term clearly dominates when $a \to 0$, and in fact renders
the average number of $i$-gons present in the ensemble with partition
function $w(z)$ {\it finite}, even at the critical point \rf{2.9}. 
This somewhat embarrassing fact can be circumvented by differentiating 
$w(z)$ a sufficient number of times with respect to  
$g$ and $z$, which will make these ``non-universal'' contributions vanish. 
For example, differentiating $w(z)$ twice with respect to $g$ yields
an expression which diverges in the limit $a\to 0$, allowing one
to ignore the finite contribution from non-universal terms,
\beq\label{2.9a}
\frac{\prt^2 w(z)}{\prt g^2} \propto \frac{1}{a^{5/2}} \; 
\frac{1}{\sqrt{\La} \;(Z+\sqrt{2\La/3})^{3/2}} \propto 
\frac{1}{a^{5/2}}\; \frac{\prt^2 W_E(Z,\La)}{\prt \La^2}.
\eeq
In this sense one may view $W_E(Z,\La)$ as the continuum disc function.
However, it would be incorrect to say that $w(z)$ was 
the regularized Hartle-Hawking
wave function from which one obtained the continuum 
Hartle-Hawking wave function
through an additive renormalization of the cosmological 
and the boundary cosmological constants.

In the following section, we will discuss a new kind of scaling limit
for matrix models, where such non-universal term are not present.

\section{The new scaling limit of matrix models}\label{sec3}

Hermitian matrix models are often analyzed in terms of the
dynamics of their eigenvalues. Since the action \rf{2.7} is invariant 
under the transformation $\phi \to U \phi U^\dg$, with $U \in U(N)$
a unitary $N\times N$-matrix, 
one can integrate out the ``angular'' degrees of freedom. What is left is
an integration over the eigenvalues $\l_i$ of $\phi$ only, 
\beq\label{3.1}
Z(g) \propto \int \prod_{i=1}^N d\l_i \; 
e^{-N \sum_j V(\l_j)} \; \prod_{k<l} |\l_k-\l_l|^2,
\eeq
where the last factor, the Vandermonde determinant, 
comes from integrating over the 
angular variables, and where
\beq\label{3.2}
{\rm tr}\ V(\phi) = \sum_{i=1}^N V(\l_i).
\eeq
Naively one might expect that 
the large-$N$ limit is dominated by a saddle-point
with $V'(\l)=0$. However, this is not the case since 
the Vandermonde determinant in \rf{3.1} contributes in the large-$N$ limit. 
The cut which appears in $w(z)$ is a direct
result of the presence of the Vandermonde determinant. In this way
one can say that the dynamics of the eigenvalues is ``non-classical'',
deviating from $V'(\l) =0$, the size of the cut being a measure of this 
non-classicality. We will now introduce a new coupling constant $g_s$ in the
matrix model by substituting
\beq\label{3.3}
V(\phi) \to \frac{1}{g_s} \; V(\phi),
\eeq
and consider the limit $g_s \to 0$. 
As will become clear in due course, this
controls and reduces the size of the cut and thus brings 
the system closer to a ``classical'' behaviour.

In the analysis it will be
convenient to keep the coupling constant $t_1 > 0$, 
which can be motivated as follows.
Consider a ``triangulation'' consisting of $T_1$ one-gons,
$T_2$ two-gons, $T_3$ triangles, $T_4$ squares etc., up to $T_n$ n-gons, and
expand the $g$-dependent part of the associated matrix integral
\beq\label{3.4}
Z(g,g_s) = \int d\phi \;\e^{-\frac{N}{g_s} \; \tr\left( \oh\phi^2 -
g \sum_{i=1}^n \frac{t_i}{i} \; \phi^i\right)}
\eeq
in powers of $g$. Each $i$-gon appearing
in the triangulation has a factor $g/g_s$ associated with it. At the same time,
the Gaussian integration will produce a factor $g_s^L$, where 
\beq\label{3.5}
L = \oh T_1 + T_2 + \frac{3}{2} T_3 +\cdots + \frac{n}{2}T_n
\eeq
is the number of links in the triangulation, since  
each Gaussian integration corresponds precisely to a gluing of an $i$-gon and
a $j$-gon.
The total coupling-constant factor associated with the 
triangulation is therefore given by
\beq\label{3.6}
g^{T_1+\cdots +T_n} g_s^{-T_1/2 +T_3/2 +\cdots + (n/2-1)T_n}.
\eeq
We observe that in the limit $g_s \to 0$, a necessary condition for obtaining 
a finite critical value $g_c(g_s)$ is $T_1 > 0$. 
We should emphasize that the analysis described below can be 
carried out also if we suppress the appearance of any one-gons
(by setting $t_1 =0$) in our triangulations, but it is 
slightly more cumbersome since then $g_c (g_s)\to \infty$ as $g_s \to 0$, 
requiring further rescalings.

For simplicity we will consider the simplest 
nontrivial model with potential\footnote{In accordance with the remark above 
this model can be related to the model without one-gons
by the field and coupling-constant redefinitions
\beq\label{3.8}
\tilde{\phi} = \sqrt{1-4g^2}\left(\phi+\frac{2g}{1+\sqrt{1-4g^2}}\right),
~~~~\tg=\frac{g}{\sqrt{1-4g^2}}.\eeq}
\beq\label{3.7}
V(\phi)= \frac{1}{g_s}\left( -g \phi + \oh \phi^2 -\frac{g}{3} \phi^3\right)
\eeq
and analyze its behaviour in the limit $g_s \to 0$.
The disc amplitude now has the form
\beq\label{3.9}
w(z) = \frac{1}{2g_s} \left(-g +z -gz^2 + g(z-b)\sqrt{(z-c)(z-d)}\right),
\eeq
and the constants $b$, $c$ and $d$ are determined by the requirement 
that $w(z) \to 1/z$ for $z \to \infty$. 
Compared with the analysis of the previous
section, the algebraic condition fixing the
coefficient of $1/z$ to be unity will now enforce a 
completely different scaling behaviour as $g_s \to 0$.

For the time being, we will think of $g_s$ as small and fixed, and perform
the scaling analysis for $g_c(g_s)$.
As already mentioned in eq.\ \rf{2.5a} above, the 
critical point $g_c$ is determined by the additional 
requirement that $b(g_c)=c(g_c)$, which presently
leads to the equation 
\beq\label{3.10}
\left(1-4g_c^2\right)^{3/2} = 12 \sqrt{3} \;g_c^2 \,g_s.
\eeq
Anticipating that we will be
interested in the limit $g_s \to 0$, we write the critical points as
\beq\label{3.11}
g_c(g_s)= \oh (1-\Del g_c(g_s)),~~~\Del g_c(g_s) = 
\frac{3}{2} g_s^{2/3} + O(g_s^{4/3}),
\eeq
and 
\beq\label{3.12}
z_c (g_s) = c(g_c,g_s)= 
\frac{1}{2g_c(g_s)}\left(1+\sqrt{\frac{1-4g_c(g_s)^2}{3}}\right) =
1 + g_s^{1/3} + O(g_s^{2/3}),
\eeq
while the size of the cut in \rf{3.7}, $c(g_c)-d(g_c)$,  behaves as 
\beq\label{3.12a}
c(g_c)-d(g_c) = 4 g_s^{1/3} + 0(g^{2/3}_s).
\eeq
Thus the cut shrinks to zero as $g_s \to 0$.

Expanding around the critical point given by \rf{3.11}-\rf{3.12} a 
nontrivial limit can be obtained if we insist that 
in the limit $a\rightarrow 0$, $g_s$ scales according to
\beq\label{3.13}
g_s = G_s a^3,
\eeq
where $a$ is the lattice cut-off introduced earlier.
With this scaling the size of the cut scales to zero as $4\,a\, G_s^{1/3}$.
In addition $\sqrt{(z-c)(z-d)} \propto a$ 
if we introduce the standard identification \rf{2.9}: 
$z=c(g_c)+a\, Z$. This scaling is different from the conventional 
scaling in Euclidean quantum gravity where $\sqrt{(z-c)(z-d)}
\propto a^{1/2}$ since in that case
$(z-c)$ scales while $(z-d)$ does not scale.

We can now write
\beq\label{3.14}
g = g_c(g_s)(1-a^2 \La) = \bg(1 - a^2 \cdtL+O(a^4)),
\eeq
with the identifications
\beq\label{3.14a}
\cdtL \equiv \La + \frac{3}{2}G_s^{2/3},~~~~\bg = \oh,
\eeq
as well as
\beq\label{3.15}
z= z_c +a Z = \bz+a \cdtZ +O(a^2),
\eeq
with the identifications
\beq\label{3.15a}
\cdtZ\equiv Z+G_s^{1/3},~~~~\bz=1.
\eeq 
Using these definitions one computes in the limit $a \to 0$ that
\beq\label{3.16}
w(z) = \frac{1}{a} \; \frac{\cdtL -\oh \cdtZ^2 + 
\oh(\cdtZ-H)\sqrt{(\cdtZ+H)^2 -\frac{4G_s}{H}}}{2G_s}.
\eeq
In \rf{3.16}, the constant $H$ (or rather, its rescaled version 
$h=H/\sqrt{2\cdtL}\;$) satisfies the third-order equation
\beq\label{3.17}
h^3 -h + \frac{2G_s}{(2\cdtL)^{3/2}} =0,
\eeq
which follows from the consistency equations for the constants $b$, $c$ and $d$
in the limit $a \to 0$.
We thus define
\beq\label{3.17a}
w(z) = \frac{1}{a}\, \cdtW(\cdtZ,\cdtL,G_s) \equiv \frac{1}{a}\, W(Z,\La,G_s)
\eeq  
in terms of the continuum 
Hartle-Hawking wave functions $\cdtW(\cdtZ,\cdtL,G_s)$ and $W(Z,\La,G_s)$.

Notice that while the cut of $\sqrt{(z-c)(z-d)}$ goes to zero as the 
lattice spacing $a$, it nevertheless survives in the scaling limit
when expressed in terms of renormalized ``continuum'' variables,
as is clear from eq.\ \rf{3.16}. Only in the limit $G_s \to 0$ it
disappears and we have 
\beq\label{3.18}
w(z) = \frac{1}{a}\, \cdtW(\cdtZ,\cdtL,G_s)~~
\underset{G_s\to 0}{\longrightarrow}~~  
\frac{1}{a} \; \frac{1}{\cdtZ +\sqrt{2\cdtL}},
\eeq  
which is the original CDT disk amplitude introduced in \cite{al}.

A number of comments are in order to put this result into 
the context of previous
work concerning the standard and generalized 2d CDT theory, as well as 
Euclidean quantum gravity:
\begin{itemize}
\item[(1)] Eqs.\ \rf{3.16} and \rf{3.17} are precisely the equations that were
derived in the generalized CDT model of references \cite{alwz,alwwz1,alwwz2}. 
This suggests that $G_s$ should be
interpreted as a coupling constant associated with
the splitting of a spatial universe into two. This is a meaningful statement 
in a universe with Lorentzian signature, 
where (in the simplest case) such a splitting 
is associated with an isolated point where the 
metric and its associated light-cone
structure are degenerate, which has a diffeomorphism-invariant meaning. 
This was the motivation for considering such processes in CDT 
in the first place, leading to a generalization of the original 
CDT disc amplitude alluded to in \rf{3.18}
to the expression $\cdtW(\cdtZ,\cdtL,G_s)$ 
given by eq.\ \rf{3.16} when $G_s > 0$. 
However, we can also formally
make such an association in this purely Euclidean matrix
model by noting that by introducing the coupling constant
$g_s$ (and assuming $t_1 = t_3 =1$ and $t_i=0$ otherwise)
eq.\ \rf{2.7a} is changed to 
\beq\label{3.18a}
w(z)= g\left(\sum_{i=1}^n t_i z^{i-2}\right) w(z) + \frac{g_s}{z}\, w^2(z)
+\frac{1}{z}\, Q(z,g).
\eeq 
Going back to Fig.\ \ref{fig2}, this suggests that one should associate
a factor $g_s$ instead of a factor 1 with the graph with the double
line. Geometrically this can be viewed as a process where a triangle is 
removed at a marked link (and a new link is marked at the new boundary),
except in the case where the marked link does not belong to a
triangle, but is part of a double-link, in which case the double link is
removed and the triangulation is separated into two. 
If one thinks of the recursion process in Fig.\ \ref{fig2}
as a ``peeling away" of the triangulation as proper time advances, the
presence of a double link represents the ``acausal" splitting point beyond
which the triangulation splits into two discs with two separate boundary
components (i.e. two separate one-dimensional spatial universes). 
Associating an explicit weight $g_s$ with this situation and  
letting $g_s \to 0$ suppresses this process compared to processes 
where we simply remove an $i$-gon from the triangulation.
Nevertheless, in the limit $a \to 0$ the
process survives precisely when we scale $g_s \to 0$ as prescribed
by eq.\ \rf{3.13}. The interpretation of this process, 
advocated in \cite{fractal}, is that it represents a split 
of the spatial boundary with respect to (Euclidean) proper time.  

\item[(2)] Using eq.\ \rf{3.17} we can expand $w(z)$ into a power series
in $G_s/\cdtL^{3/2}$ whose radius of convergence is $1/3\sqrt{3}$. 
For fixed values of $\cdtL$, this value corresponds to the largest 
value of $G_s$ where \rf{3.17} has a 
positive solution for $h$. 
The existence of such a bound on $G_s$ for fixed $\cdtL$
was already observed in \cite{alwz}.
This bound can be re-expressed more transparently in the present
Euclidean context, where it is more natural to keep the ``Euclidean"
cosmological constant $\La$ fixed, rather than
$\cdtL$. We have
\beq\label{3.19}
\frac{G_s}{(2\cdtL)^{3/2}}\leq  \frac{1}{3\sqrt{3}}~ \Rightarrow ~ 
\frac{{3} G_s^{2/3}}{2\La +{3} G_s^{2/3}} \leq 1,
\eeq
which for fixed $\La > 0$ is obviously satisfied for all positive
$G_s$. In order to see that the usual Euclidean 2d quantum gravity 
(characterized by some {\it finite} value for $g_s$) can be rederived
from the disc amplitude \rf{3.16}, let us 
expand eq.\ \rf{3.16} for large $G_s$.
The square root part becomes
\beq\label{3.20}
a^{-1}\;G_s^{-5/6}\; 
\left(Z- \sqrt{2\La/3}\right)\sqrt{Z+2\sqrt{2\La/3}} , 
 \eeq
which coincides with the generic expression $a^{3/2} W_E(Z,\La)$  
in Euclidean 2d quantum gravity (c.f. eqs.\ \rf{2.6a} and \rf{2.6b})
if we take $G_s$ to infinity as $g_s/a^3$. However, if we reintroduce
the same scaling in the $V'(z)$-part of $w(z)$, it does not scale with $a$
but simply goes to a constant. This term would dominate $w(z)$
in the limit $a \to 0$ if one did not remove it by hand, as is usually done
in the Euclidean model (see Sec.\ \ref{sec2}).

\item[(3)] Why does the potential $V'(z)$ (and therefore the entire
disc amplitude $w(z)$) scale (like $1/a$) in the new continuum limit with 
$g_s = G_s a^3$, $a\to 0$, contrary
to the situation in ordinary Euclidean quantum gravity?
This is most clearly seen by looking again at 
the definitions \rf{3.14} and \rf{3.15}.
Because of the vanishing
\beq\label{3.21}
V'(\bz,\bg)=0,~~~~V''(\bz,\bg)=0
\eeq
in the point $(\bz,\bg)=(1,1/2)$,
expanding around $(\bz,\bg)$ according to \rf{3.14}, \rf{3.15}
leads automatically to a potential which 
is of order $a^2$ when expressed in terms
of the renormalized constants $(\cdtZ,\cdtL)$,
precisely like the square-root term when expressed in terms of 
$(\cdtZ,\cdtL)$. 

The point $(\bz,\bg )$ differs from the critical point 
$(z_c(g_s),g_c(g_s))$, as
long as $g_s\not= 0$.
In fact, both $1/\bz$ and $\bg$ lie {\it beyond} 
the radii of convergence of 
$1/z$ and $g$, which are precisely $1/z_c(g_s)$ and $g_c(g_s)$. 
However, since the differences
are of order $a$ and $a^2$, respectively, they simply amount to {\it shifts}
in the renormalized variables, as
made explicit in eqs.\ \rf{3.14a} and \rf{3.15a}. Therefore, re-expressing 
$W(Z,\La,G_s)$ in \rf{3.17a} in terms of the variables 
$\cdtZ$ and $\cdtL$ simply leads to the expression 
$\cdtW(\cdtZ,\cdtL,G_s)$, first derived in \cite{alwz}.
Similarly, any geometric quantities defined with respect to $Z$ and $\La$ 
can equally well be expressed in terms of $\cdtZ$ and $\cdtL$.
For instance, the average continuum length of the boundary and the average 
continuum area of a triangulation are given by
\beq\label{3.22}
\la L \ra = \frac{\prt \ln W(Z,\La,G_s)}{\prt Z}= 
\frac{\prt \ln \cdtW(\cdtZ,\cdtL,G_s)}{\prt \cdtZ},
\eeq
\beq\label{3.23}
\la A \ra =\frac{\prt \ln W(Z,\La,G_s)}{\prt \La}= 
\frac{\prt \ln \cdtW(\cdtZ,\cdtL,G_s)}{\prt \cdtL}. 
\eeq
In the limit of $G_s \to 0$,
the variables $(Z,\La)$ and $(\cdtZ,\cdtL)$ become identical and the 
disc amplitude becomes the original CDT amplitude 
$(\cdtZ+\sqrt{2\cdtL})^{-1}$ alluded to in \rf{3.18}.

\item[(4)] In the matrix potential \rf{3.7}, which formed the starting point
of our new scaling analysis, we are still free to perform a change 
of variables. Inspired by
relations \rf{3.14}--\rf{3.15a}, let us transform to new ``CDT''-variables
\beq\label{3.24}
\phi \to \bz \, \hI + a \Phi + O(a^2),
\eeq 
at the same time re-expressing $g$ as
\beq
g=\bg(1-a^2\cdtL+O(a^4)),
\eeq
following eq.\ \rf{3.14}.
Substituting the variable change into the matrix potential, and discarding
a $\phi$-independent constant term, one obtains
\beq\label{3.25}
V(\phi) = \bV(\Phi) \equiv \frac{\cdtL \Phi -\frac{1}{6} \Phi^3}{2G_s}
\eeq
in the limit $a \to 0$, from which it follows that
\beq\label{3.26}
Z(g,g_s) = a^{N^2} Z(\cdtL,G_s), ~~~Z(\cdtL,G_s) = \int d\Phi \; 
\e^{-N\tr \bV(\Phi)}.
\eeq
The disc amplitude for the potential $\bV(\Phi)$ is precisely 
$W(\cdtZ,\cdtL,G_s)$, and since by definition 
\beq\label{3.27}
\frac{1}{z-\phi}= \frac{1}{a}\, \frac{1}{\cdtZ-\Phi},
\eeq
the first equal sign in eq.\ \rf{3.17a} follows straightforwardly from
the simple algebraic equation \rf{3.25}. 
We conclude that  the continuum generalized
CDT-theory is described by the matrix model with potential $\bV(\Phi)$.
While in the present article
we have analyzed this only to leading order in $N$,
the proof that the generalized CDT-theory is reproduced
by the matrix model with potential \rf{3.25} to all orders in $1/N^2$ 
was already given in \cite{alwwz2}.
The important new point made here is that there really is a 
regularized lattice theory with a geometric
interpretation underlying the
formal manipulations of eqs.\ \rf{3.25}-\rf{3.26}, 
where $a$ appears merely as a parameter without an obvious 
geometric interpretation as a cut-off.

\item[(5)] We can generalize the discussion above if we drop the 
restriction that all $t_i$ should be positive. Then there exist choices 
such that 
\beq\label{3.28}
V'(\bz,\bg,\bt_i)= \cdots = V^{(n)}(\bz,\bg,\bt_i) =0,
\eeq
where the derivatives are with respect to $z$. For a potential 
satisfying \rf{3.28} we obtain
a nontrivial limit if we introduce the scaling
\beq\label{3.29}
g_s = G_s a^n,~~~g=\bg(1- a^n\cdtL^{n/2}),~~~z=\bz(1-a\cdtZ).
\eeq
In the limit $a\to 0$ one obtains
\beq\label{3.30}
\cdtW(\cdtZ,\cdtL) = \frac{V_n(\cdtZ)
+  P_n(\cdtZ)\sqrt{(\cdtZ-C)(\cdtZ-D)}}{2G_s},
\eeq
\beq\label{3.30a}
V_n(\cdtZ) \equiv \cdtL^{n/2}-
\frac{(-1)^n}{n}\cdtZ^n,
\eeq
where $P_n(\cdtZ)$ is a polynomial of order $n-1$. The coefficients
of $P_n(\cdtZ)$ as well as $C$ and $D$ satisfy algebraic equations
which are derived from the requirement that $\cdtW(\cdtZ)$ fall off like
$1/\cdtZ$. In the limit $G_s\to 0$ one finds
\beq\label{3.31}
\cdtW(\cdtZ,\cdtL,G_s) \to \frac{1}{\cdtZ+ \sqrt{2\cdtL}},
\eeq
which coincides with the original CDT result, valid when no splitting of the
one-dimensional spatial universe is allowed as a function of proper time.

The scaling of $g$ in \rf{3.29} is identical to the scaling
for the conventional multicritical one-matrix model 
\cite{kazakov,staudacher} (for reviews see \cite{gm,difgz}). 
Precisely as in that case one has a $(n-1)$-dimensional set of 
deformations around the solution \rf{3.30}.

\end{itemize}

\section{Discussion}\label{discuss}

The CDT model  has been advocated as a potential candidate
for quantum gravity in four dimensions \cite{agjl,ajl}, the triangulations
in this case constructed by gluing together four-simplices in such a 
way that the geometry is causal in the sense defined in GR. If one 
does not impose any causal constraint these regularized models have so
far not led to a continuum theory in four dimensions 
\cite{bielefeld}. It is of interest to study to what extent one
can lift the causality constraint in the CDT model and study the relation
to the pure Euclidean model. This was one of the main motivations
for the present investigation in two dimensions. The two-dimensional
models are of course only toy models of quantum gravity, but they
have the advantage that explicit analytic calculations can be performed.  
In this spirit it was shown in \cite{alwwz2} that the loop equations
of the matrix model with the cubic potential \rf{3.25}
are identical to the continuum string-field equations of a 
generalized CDT model. The generalization consisted in the inclusion of
topology changes of the spatial universe, which because of their
causality-violating nature were excluded from the original 2d CDT model.
However, it left open the issue of
how to understand this generalized continuum model
as the scaling limit of an ordinary matrix model, with the conventional
geometric interpretation of implementing a gluing of 
polygons of side length $a$, $a$ being the ultraviolet lattice
cut-off, and in addition to understand
the precise relation of this model to ``ordinary'' two-dimensional 
Euclidean quantum gravity. 
In this article we have provided such an interpretation and worked
out in detail the relation to Euclidean quantum gravity.

This enables us to view the matrix models with a potential of the 
type \rf{2.7} as representing regularized path integrals over 
random surfaces, as one 
would like to do in a theory of two-dimensional quantum gravity. 
What is rather remarkable is that starting with 
such a matrix model and taking the 
continuum limit $a \to 0$, while scaling 
$g_s \to 0$ as $G_s a^3$, one again ends up
with a matrix model, namely, the one given by \rf{3.25}, 
now describing the {\it continuum} CDT string field theory
formulated in \cite{alwwz1}. 

In the CDT framework, the new coupling 
constant $G_s$ describes the splitting of a spatial 
hypersurface into two. Such a concept is not topological in a two-dimensional
sense, but has 
a diffeomorphism-invariant meaning in spacetimes of Lorentzian 
signature. Well-behaved causal properties of spacetime were 
the starting point of the original CDT path-integral formulation, motivated
by arguments going back to \cite{teitelboim}. Each individual, causal 
configuration in the regularized CDT path integral allowed
a rotation to a unique Euclidean spacetime, leading to
a modified Euclidean path integral with many fewer geometries than 
the ordinary Euclidean gravitational path integral. What we have
shown here is that also this restricted class of geometries can be
captured by starting out with an ordinary matrix model (i.e. one intended 
to describe Euclidean 2d quantum gravity), and then imposing a ``penalty"
in the form of a coupling constant $g_s = G_s a^3$
for the process where the boundary splits into two. 

   
A potentially powerful application of the model described here concerns
matter-coupled quantum gravity.
It has been shown that in the continuum limit 
the triangulations appearing in CDT are much more
regular than generic triangulations of the disc: a
generic triangulation of the disc has fractal dimension four 
\cite{fractal}, while those appearing in CDT have (fractal) dimension
two \cite{alwz,fractal2}. As a result,
when one couples the Ising model to two-dimensional Euclidean quantum gravity
the critical matter exponents will change from the Onsager values to the 
so-called KPZ values. This change can be traced to the fractal
nature of the generic triangulations on which the Ising spins are placed,
using DT as the framework for (regularized) two-dimensional Euclidean
quantum gravity. By contrast, if one couples the Ising model to Lorentzian
quantum gravity in the form of CDT triangulations,
one obtains the Onsager values for the critical exponents 
as has been shown numerically \cite{aal} and using a high-$T$ expansion
\cite{bl}. By generalizing the
framework of this article from a one-matrix 
to a two-matrix model description, one may be able to obtain 
a two-matrix model description of CDT coupled to the Ising model and
in this way prove analytically that the critical exponents 
of the Ising model on CDT are indeed the Onsager exponents. If this can be
done, it might be by far the simplest way to calculate explicitly the 
Onsager exponents of an Ising lattice model. Work in this direction will 
be reported elsewhere.    

It is also somewhat surprising that the matrix integral \rf{3.26}, 
with the {\it same} dimensions associated
to the fields and coupling constants, is encountered
in the so-called Dijkgraaf-Vafa correspondence \cite{dv}.
From the gauge theory side  $V(\Phi)$ is then 
the tree-level superpotential of the adjoint 
chiral field $\Phi$, which breaks the 
supersymmetry of the unitary gauge theory from $\cN = 2$ to $\cN = 1$.   
If one demands that this tree-level potential corresponds to a 
renormalizable theory then the form of the potential $V(\Phi)$ is
essential unique (and given by \rf{3.25}). 
$G_s$ is a dimension three coupling constant coming either from
three of the compactified dimensions in  
topological string theory or alternatively, via the DV-correspondence,
from the glueball superfield condensate in the gauge theory.
In this formalism a continuum interpretation is associated 
with the equation defining disk amplitude, rather than 
with the (discretized) surfaces defined by the matrix integral, the
equation defining an algebraic surface. This point of view is 
also adopted in \cite{eynard,ce,eo1} where the calculation
of matrix integrals like \rf{3.26} has been carried to a new level
of perfection. In these applications one encounters eventually much 
more complicated situations with higher order potentials $V(\Phi)$ 
which admit multi-cut solutions. It would be interesting if these 
multi-cut solutions could also find a random surface interpretation.

\section*{Acknowledgment}
JA, RL, WW and SZ acknowledge the  support by
ENRAGE (European Network on
Random Geometry), a Marie Curie Research Training Network in the
European Community's Sixth Framework Programme, network contract
MRTN-CT-2004-005616. RL acknowledges
support by the Netherlands
Organisation for Scientific Research (NWO) under their VICI
program.


\end{document}